\let\includefigures=\iftrue

\input harvmac
\newcount\yearltd\yearltd=\year\advance\yearltd by 0



\input epsf

\newcount\figno
\figno=0
\def\fig#1#2#3{
\par\begingroup\parindent=0pt\leftskip=1cm\rightskip=1cm\parindent=0pt
\baselineskip=11pt
\global\advance\figno by 1
\midinsert
\epsfxsize=#3
\centerline{\epsfbox{#2}}
\vskip 12pt
{\bf Figure \the\figno:} #1\par
\endinsert\endgroup\par
}
\def\figlabel#1{\xdef#1{\the\figno}}


\noblackbox
\def\IZ{\relax\ifmmode\mathchoice
{\hbox{\cmss Z\kern-.4em Z}}{\hbox{\cmss Z\kern-.4em Z}}
{\lower.9pt\hbox{\cmsss Z\kern-.4em Z}} {\lower1.2pt\hbox{\cmsss
Z\kern-.4em Z}}\else{\cmss Z\kern-.4em Z}\fi}

\font\cmss=cmss10 \font\cmsss=cmss10 at 7pt
\def\IR{\relax{\rm I\kern-.18em R}}

\def\frac#1#2{{#1 \over #2}}


\def\frac#1#2{{#1\over#2}}

%
%

\overfullrule=0pt

\def\underrel#1\over#2{\mathrel{\mathop{\kern\z@#1}\limits_{#2}}}


%

\lref\tomcsm{
T. Banks,
``Cosmological Breaking of Supersymmetry ? Or Little Lambda Goes Back
to the Future 2.'',
hep-th/0007146.}

\lref\polbou{ J. Brown and C. Teitelboim, Phys. Lett. B195 177
(1987); Nucl. Phys. B297, 787 (1988);
R. Bousso and J. Polchinski, ``Quantization of Four-form Fluxes and
Dynamical Neutralization of the Cosmological Constant'',
hep-th/0004134, JHEP 0006 (2000) 006;
J.L. Feng, J. March-Russell, S. Sethi and F. Wilczek, ``Saltatory
Relaxation of the Cosmological Constant'', hep-th/0005276,
Nucl. Phys. B602 (2001) 307-328. }

\lref\edcosm{
E. Witten,
``Strong Coupling and the Cosmological Constant'',
hep-th/9506101,
Mod.Phys.Lett. A10 (1995) 2153-2156.
}

\lref\noncrit{A. Maloney, E. Silverstein and A. Strominger
``de Sitter Space in Non-Critical String Theory'',
hep-th/0205316, Hawking Festschrift.
}

\lref\twoloop{
S. Kachru and E. Silverstein,
``On Vanishing Two Loop Cosmological Constant in Nonsupersymmetric Strings'',
hep-th/9810129, JHEP 9901 (1999) 004.
}

\lref\slftun{
 S. Kachru, M. Schulz and E. Silverstein,
``Self-tuning flat domain walls in 5d gravity and string theory'',
hep-th/0001206,
Phys.Rev. D62 (2000) 045021.
}

\lref\infex{
G. Dvali, G. Gabadadze and M. Shifman,
``Diluting Cosmological Constant In Infinite Volume Extra'',
hep-th/0202174.
}

\lref\dvalib{ G. Dvali, G. Gabadadze and M. Shifman,
``Diluting Cosmological Constant via Large Distance Modification of Gravity''
hep-th/0208096.
}

\lref\grg{
G.W. Moore,
``Atkin-Lehner Symmetry'',
Nucl. Phys. B293:139, 1987, Erratum-ibid. B299:847, 1988.
}

\lref\rubashap{
V.A. Rubakov and M.E. Shaposhnikov,
``Extra Space-time Dimensions: Towards a Solution to the Cosmological
Constant Problem'',
Phys.Lett.B125:139,1983.
}

\lref\akama{
By K. Akama, ``Pregeometry'', hep-th/0001113, Lect. Notes
Phys.176:267-271,1982 Also in *Nara 1982,
Proceedings, Gauge Theory and Gravitation*, 267-271.
}

\lref\rstwo{
L. Randall and R. Sundrum,
``An Alternative to Compactification'', hep-th/9906064,
Phys.Rev.Lett.83:4690-4693,1999.
}

\lref\visser{
M. Visser,
``An Exotic Class of Kaluza-Klein Models'', hep-th/9910093,
Phys. Lett. B159:22, 1985.
}

\lref\malda{
J.~Maldacena,
``The large $N$ limit of superconformal field theories and
supergravity,'' hep-th/9711200,
Adv.\ Theor.\ Math.\ Phys.\ 2, 231 (1998), Int.\ J.\ Theor.\ Phys.\ 38, 1113 (1998).
}

\lref\omeb{
O. Aharony, M. Berkooz and E. Silverstein,
``Non-local String Theories on $AdS_3\times S^3$ and
non-supersymmetric backgrounds'',
hep-th/0112178, Phys. Rev. D65 (2002) 106007.
}

\lref\omea{
O. Aharony, M. Berkooz and E. Silverstein,
``Multiple-Trace Operators and Non-Local String Theories'',
hep-th/0105309,
JHEP 0108 (2001) 006.}

\lref\bss{
M. Berkooz, A. Sever and A. Shomer, ``Double-trace
Deformations, Boundary Conditions and Space-time Singularities'',
hep-th/0112264, JHEP 0205 (2002) 034.
}

\lref\wtnbndry{
E. Witten,
``Multi-Trace Operators, Boundary Conditions, And AdS/CFT Correspondence'',
hep-th/0112258.
}

\lref\jss{
A. Sever and A. Shomer,
``A Note on Multi-trace Deformations and AdS/CFT'',
hep-th/0203168, JHEP 0207 (2002) 027.
}

\lref\vijay{V. Balasubramanian, P. Kraus and A. Lawrence
``Bulk vs. Boundary Dynamics in Anti-de Sitter Space-time''
hep-th/9805171, Phys.Rev. D59 (1999) 046003.}

\lref\mlstlong{
J. Maldacena, J. Michelson and A. Strominger,
``Anti-de Sitter Fragmentation'',
hep-th/9812073, JHEP 9902 (1999) 011.
}

\lref\sewtlong{
N. Seiberg and E. Witten,
``The D1/D5 System And Singular CFT'',
hep-th/9903224, JHEP 9904 (1999) 017.
}

\lref\maoga{J. Maldacena and H. Ooguri,
``Strings in $AdS_3$ and the $SL(2,R)$ WZW Model. Part 1: The Spectrum'',
hep-th/0001053,
J .Math.Phys. 42 (2001) 2929-2960.
}

\lref\linde{R. Kallosh, A.D. Linde, S. Prokushkin and M. Shmakova,
``Supergravity, Dark Energy and the Fate of the Universe'',
hep-th/0208156.}

\lref\hullnc{ C.M. Hull and N.P. Warner, ``Non-compact Gauging
from Higher Dimensions'', Class. Quant. Grav. 5,1517, 1988.}

\lref\rena{R. Kallosh, A.D. Linde, S. Prokushkin and M. Shmakova,
``Gauged Supergravities, de Sitter space and Cosmology'',
hep-th/0110089, Phys. Rev. D 65, 105016 (2002).}

\lref\renb{R. Kallosh,
``Supergravity, M-Theory and Cosmology'',
hep-th/0205315.}

\lref\gpnr{D. Gepner,
``Lectures on N=2 String Theory'',
In Superstrings 89, The Trieste Spring School, 1989.
}

\lref\emila{
E.J. Martinec and W. McElgin,
``String Theory on AdS Orbifolds''
hep-th/0106171, JHEP 0204 (2002) 029.
}

\lref\emilb{E.J. Martinec and W. McElgin,
``Exciting AdS Orbifolds'',
hep-th/0206175.
}

\lref\vijcon{
V. Balasubramanian, J. de Boer, E. Keski-Vakkuri and S.F. Ross,
``Supersymmetric Conical Defects'',
hep-th/0011217,
Phys.Rev. D64 (2001) 064011.
}

\lref\susycond{
A, Giveon and M. Rocek,
``Supersymmetric String Vacua on $AdS_3\times {\cal N}$'',
hep-th/9904024.}

\lref\GiveonNS{
A.~Giveon, D.~Kutasov and N.~Seiberg,
``Comments on string theory on $AdS_3$,''
he-th/9806194,
Adv. Theor. Math. Phys. 2, 733 (1998).
}

\lref\wtndl{
E.~Witten,
``Anti-de Sitter space and holography,'' hep-th/9802150,
Adv. Theor. Math. Phys. 2, 253 (1998).
}

\lref\kgp{
S.~S.~Gubser, I.~R.~Klebanov and A.~M.~Polyakov,
``Gauge theory correlators from non-critical string theory,'' hep-th/980210,
Phys. Lett. B 428, 105 (1998).
}

\lref\adsrev{O. Aharony, S.S. Gubser, J. Maldacena, H. Ooguri and
Y. Oz,
``Large N Field Theories, String Theory and Gravity'',
hep-th/9905111, Phys.Rept. 323 (2000) 183-386.
}

\lref\misj{
M. Berkooz and S.-J. Rey,
``Non-Supersymmetric Stable Vacua of M-Theory'',
hep-th/9807200,
JHEP 9901 (1999) 014.
}

\lref\rob{
R.G. Leigh,
``Dirac-Born-Infeld Action from Dirichlet Sigma Model'',
Mod. Phys. Lett.A4:2767,1989.}

\lref\bgrglp{
J. Bagger and A. Galperin
``Linear and Non-linear Supersymmetries'',
hep-th/9810109,
*Dubna 1997, Supersymmetries and quantum symmetries* 3-20.}

\lref\kw{I.~R.~Klebanov and E.~Witten,
``AdS/CFT correspondence and symmetry breaking,'' hep-th/9905104,
Nucl. Phys. B 556 (1999) 89.}

\lref\evaallan{
A. Adams and E. Silverstein,
``Closed String Tachyons, AdS/CFT and Large N QCD'', hep-th/0103220,
Phys. Rev. D64:086001, 2001.}

\lref\kraus{
P. Kraus and E.T. Tomboulis
``Title: Photons and Gravitons as Goldstone Bosons, and the Cosmological Constant'',
hep-th/0203221, Phys.Rev. D66 (2002) 045015.
}

\lref\hrmn{H. Verlinde,
``Holography and Compactification'',
hep-th/9906182, Nucl. Phys. B580 (2000) 264-274.
}

\lref\gkp{S.B. Giddings, S. Kachru and J. Polchinski,
``Hierarchies from Fluxes in String Compactifications'',
hep-th/0105097.}

\lref\evarec{A. Adams, J. McGreevy and E. Silverstein,
"Decapitating Tadpoles", hep-th/0209226.}

\lref\savrec{N. Arkani-Hamed, S. Dimopoulos, G. Dvali and G.
Gabadadze, "Non-Local Modification of Gravity and the Cosmological
Constant Problem", hep-th/0209227.}

\noblackbox



\def\myTitle#1#2{\nopagenumbers\abstractfont\hsize=\hstitle\rightline{#1}%
\vskip 0.5in\centerline{\titlefont #2}\abstractfont\vskip .5in\pageno=0}

\myTitle{\vbox{\baselineskip12pt\hbox{hep-th/0209257}
\hbox{WIS/41/02-SEPT-DPP}
}} {\vbox{
        \centerline{Double Trace Deformations, Infinite Extra Dimensions,}
        \medskip
        \centerline{and Supersymmetry Breaking}}}
\medskip
\centerline{Micha Berkooz\foot{
Incumbent of the Recanati
career development chair of energy research. E-mail :
{\tt Micha.Berkooz@weizmann.ac.il.}}
}
\medskip
\centerline{Department of Particle Physics, The Weizmann
Institute of Science,}
\centerline{Rehovot 76100, Israel}

\bigskip
\noindent

It was recently shown how to break supersymmetry in certain
$AdS_3$ spaces, without destabilizing the background, by using a
``double trace'' deformation which localizes on the boundary of
space-time. By viewing spatial sections of $AdS_3$ as a
compactification space, one can convert this into a SUSY breaking
mechanism which exists uniformly throughout a large 3+1
dimensional space-time, without generating any dangerous tadpoles.
This is a generalization of a Visser type infinite extra
dimensions compactification. Although the model is not Lorentz
invariant, the dispersion relation is relativistic at high enough
momenta, and it can be arranged such that at the same kinematical
regime the energy difference between between former members of a
SUSY multiplet is large.

\Date{September 2002}

\baselineskip=16pt

\newsec{Introduction}

One of the main puzzles facing String theory is that of the
cosmological constant. Although several possible approaches have
been proposed (\tomcsm, \polbou, \edcosm, \noncrit, \twoloop,
\slftun, \rubashap, \infex, \grg and others), no calculable and
natural idea has been demonstrated convincingly (at least yet, as
some of the approaches are currently being investigated). The
problem appears in vacua without supersymmetry, in which one often
runs into the related problem of moduli stabilization. It may be
that we just did not find yet the correct vacuum of String theory,
or it may be that we are missing a key conceptual piece of the
puzzle. It is therefore worth exploring new mechanisms that might
be relevant to this puzzle, even if at their preliminary stages
they are not phenomenologically viable.

One class of attempts to solve the cosmological constant problem
relies on infinite extra dimensions compactifications
\rubashap\akama\visser\rstwo. In this paper we will discuss such a
compactification based on $AdS_3$, and explore its relation to double
trace deformations\foot{For previous work on double trace operators in
the context of the AdS/CFT correspondence see for example \evaallan.}
\omea\ and SUSY breaking \omeb. This will give a new way in which
infinite extra dimensions might help solve the cosmological constant
problem. Double trace deformations can be used to change the boundary
action and boundary conditions \bss\wtnbndry\jss\ on fields in the
non-compact directions transverse to our world, and SUSY breaking will
occur when we will introduce SUSY breaking boundary
conditions. However, only in very special spaces will the change of
boundary conditions influence the physics, and we will focus on one
such case.

The example will be (primarily) the familiar $AdS_3\times S^3
\times T^4$ background of string theory, which is dual
\malda\wtndl\kgp\ (for a review see \adsrev) to a 1+1 conformal
field theory on a moduli space of Instantons.  To make this into
an infinite extra dimension compactification we will pick a $T^3$
out of the $T^4$, and take it to be of very large radius relative
to the radius of curvature of $AdS_3$ (which we will keep finite,
although smaller than the string scale). The remaining circle from
the $T^4$ will be taken to be small. The time direction of the
$AdS_3$ and the 3 large coordinates of the $T^3$ can then be
considered as large dimensions, and the rest as ``compactification
dimensions''. The modes that we will be interested in are the
normalizable modes on $AdS_3$. In terms of relation to previous
work on large extra dimension, this model is a generalization of
\visser.

Our main interest will be in the new SUSY breaking mechanism
described in \omeb, which uses a double trace deformation \omea\
of the CFT to break SUSY and can be shown to not destabilize the
vacuum, even at the non-perturbative level. In the low energy
effective action, multi trace deformations localize to the
boundary of space \bss\wtnbndry\jss, and since the boundary of
space-time is parallel to $R^1\times T^3$, we get uniform SUSY
breaking throughout space-time. Generally speaking multi-trace
deformations permit the introduction of new parameters into String
theory, which can be used to change the effective action.  But
contrary to one of the dogmas of String theory, these parameters
are not given by expectation values of scalar fields, alleviating
the generic Dine-Seiberg problem of dynamical SUSY breaking in
String theory.

The outline of the paper is the following. In section 2 we
summarize the model and discuss some algebraic aspects of infinite
extra dimension compactification. The main new results of the
paper are in sections 3 and 4. In section 3 we discuss
supersymmetry breaking. We start with a supersymmetric model and
deform it to a new theory where for a class of particles, in a
kinematical regime where their dispersion relation is
approximately relativistic, the splitting between between former
members of a SUSY multiplet is large. In section 4 we point out,
based on some known facts about quantization of $SL(2,R)$, some
novel features of the 3+1 dimensional effective action.   In
section 5 we discuss generalizations of the model, which might
help remedy some of the phenomenological difficulties.

\newsec{Summary of the Model}

\subsec{The Model}

The basic example that we will explore is $AdS_3\times S^3 \times
T^4$. We will single out 3 circles out of the $T^4$ and take them
to be very large. We will view these 3 circles + the time
direction in global $AdS_3$ as our large space and we will
consider the rest as a compactification. The remaining circle of
the $T^4$ will be taken to be small.

We will denote the coordinates of the large $T^3$ inside $T^4$ by
$x^1..x^3$, and its volume by $v_3$. The 4th circle will be denoted
by $x^4$ and will satisfy $x^4\sim x^4+R_4$. The relevant formulas for
the background are the (following the notations of \GiveonNS)
\eqn\adstr{ {1\over g_s^2}={p\over
v_3R_4 k} }
$$ds^2= k{r^2\over l_s^2}d\gamma
d{\bar\gamma} + kl_s^2 ({1\over r^2} dr^2 + d\Omega_3^2) + dx_idx^i +
dx_4dx^4 $$
$$H_0=2k(\epsilon_3+*_6\epsilon_3), $$ where $g_s$ is the 10D string
coupling, $k$ is the level of the $SL(2,R)$ and of the
$SU(2)_{WZW}$, $p$ is an positive integer, $H_0$ is the NS-NS
3-form field strength, $\epsilon$ is the volume form on $AdS_3$,
and $*_6\epsilon$ is the volume form on the $S^3$. The background
is the near horizon limit $p$ fundamental strings and $k$ NS
5-branes.

The regime of parameters that we are interested in is
\item{1.}  \eqn\kscl{k>>1} In this paper we
would like to have a gap between the ``Long string modes''
(\mlstlong\sewtlong\maoga) and the gravity fields, since we would like
to concentrate on the discrete spectrum. It may be that one can relax
this condition, since the CFT at finite $k$ is under some control.
\item{2.}  \eqn\vscl{v_3/l_s^3>>\sqrt{k}} This is a condition that the
momentum modes in the $T^3$ direction will be more finely spaced than the
discrete modes in the spatial slices of $AdS_3$, such that we will
view the $T^3$ as the large dimensions, and spatial slices of $AdS_3$
(and the $S^3$) as a compactification.
\item{3.}  \eqn\gscl{g_s<<1}
\item{4.}  \eqn\rscl{R_4\sim l_s} such that we can regard the 4th
circle of the $T^4$ as a compactification.

Condition 2 is to be taken with a grain of salt since the
compactification is not Lorentz invariant. It guarantees that momentum
modes in the $T^3$ are more finely spaced than in the $AdS$ direction,
but all of this is on top of energies whose scale is set by the $AdS$
scale. One can either accept this violation of Lorentz invariance, or
we can discuss momenta larger than the $AdS$ scale. In this case there
would be many modes below our kinematical scale, but this is no worse
than most infinite extra dimension compactifcations. We will use this
kinematical regime in section 3 when discussing SUSY breaking, and
we will comment in section 5 on how one might proceed to improve
Lorentz invariance.

The fields that we are interested in - the ``stuff'' that makes
out matter in our world - are the normalizable modes of global
$AdS_3$. These have a discrete physical spectrum with the
dispersion relation\foot{The choice of sign before the $\sqrt{}$
can be changed in some cases \kw. In fact it has to be changed in
some SUSY cases. We will focus on the simpler case here, which is
valid for large enough $m$.} \eqn\disprel{w = \Lambda(1+n_L+n_R)+
\sqrt{\Lambda^2+m^2+q^2}} where $m$ is the 6 dimensional mass of
the particle on $AdS_3\times T^3$, $q$ is the 3D momentum (in
$T^3$) and $\Lambda$ is the scale set by $AdS_3$. As happens in
many infinite extra dimension compactifications\foot{and in some
supergravity models that have a $dS$ vacuum \rena\renb} the
spectrum is spaced with some spacing set by a cosmological
parameter \linde\hullnc\ (in other infinite extra dimension
compactification it is a continuum).

One may complain that the choice of a time direction in $AdS_3$ in
not unique because of the conformal invariance of $AdS$. But a 3+1
dimensional observer will report the same fact as the result of a
spectrum generating algebra (the space-time conformal symmetry of
the model). Alternatively, in more general cases, one can try to
break conformal invariance by a relevant operator. In this case
there might be a preferred time direction (and the spectrum
generating algebra would only be asymptotic at high energies).

\subsec{Properties of the Model}

Let us summarize some properties of the model (some were already
mentioned before):

\medskip

\item{1.} The space has a translationally invariant large
$R^{1,3}$ component. At length scales much smaller than the radii
of $T^3$, it also has an approximate $SO(3)$ symmetry.

\medskip

\item{2.} The 3+1 theory has a spectrum generating algebra, which
is the space-time conformal symmetry, and a known dual which
allows us non-perturbative control over the dynamics.

\medskip

\item{3.} Novel features of the action: as discussed before, it
has proven difficult to address the cosmological constant within
the ordinary rules of low energy effective action (i.e a
Lagrangian with up to 2 derivatives, analytic in the fields and
their derivatives at low momentum). It is therefore interesting to
explore what more exotic effective actions can appear in String
theory. We will see in section 4 that we obtain a rather peculiar
effective action already at the level of the free action.

\medskip

\item{4.} The background is not Lorentz Invariant: The dispersion
relation is given in \disprel. However, we can go to high momentum
relative to $\Lambda$ where it would appear to be relativistic again
(but in this regime we probe the higher dimensional space). We will
see, however, that this regime is interesting for the purposes of SUSY
breaking. We will also comment on how to try to improve the situation
in section 5.

\item{} More generally, given that we have not been very
successful in explaining the cosmological constant so far, perhaps
it is worthwhile giving up some current principles such as Lorentz
invariance at ultra low energies\foot{Another approach to the
cosmological constant problem using Lorentz symmetry breaking
appears in \kraus}. For example, in the case we are discussing
here, the non Lorentz invariant dispersion relation \disprel\
removes the zero energy mode of any scalar field. It therefore
restricts the extent to which quantum corrections can destabilize
the vacuum.

\medskip

\item{5.} A new SUSY breaking mechanism, discussed in section 3,
which does not destabilize the vacuum: even though the physical
normalizable modes on $AdS_3$ do not contain a zero energy mode
for any field, it is still rather difficult to find a
non-supersymmetric stable $AdS_3\times M$ background. The reason
is that scalar fields that correspond to marginal operators in the
dual theory can develop tadpoles, uniformly throughout $AdS$, that
cannot be compensated in a way that preserves conformal invariance
and they typically destabilize the vacuum \misj.

\medskip

\item{6.} The spectrum does not posses any zero energy
states\foot{A different way of modifying the low energy dispersion
relation in String theory was recently suggested in \evarec, and
in low energy effective action in \savrec. The relation between
these discussions and the mechanism discussed in this paper
remains to be elaborated} (in particular, a massless graviton):
Since the energy eigenvalues are related to the dimension of the
operator on the dual theory, it is clear that we cannot have any
zeromodes in this theory - this would correspond to a dimension 0
operator (A compactification with no physical zero energy modes is
interesting in itself).

\item{} It is worth noting two things. The first is that the sizes
of the $T^3$ are moduli of the CFT. Hence, they correspond to mass
zero in formula \disprel. This implies that there are low energy
symmetric 2-tensor fields. Another point is that modes of the
gravitons with polarization in directions of the $T^3$ are scalar
operators from the point of view of the dual CFT. Since the only
bound on the dimension of scalar operators in 1+1 dimensions is
that it would be larger than zero (both in the non-supersymmetric
and the supersymmetric cases), it is perhaps possible to find a
generalization of the model which will contain a graviton with
very low energy modes.

\subsec{Relation to previous work}

This compactification might be considered as an ``infinite extra
dimension'' compactification since the proper distance along a spatial
section (t=const) of the $AdS_3$ from its center to the boundary is
infinite. Despite this fact, one obtains a discrete on-shell spectrum,
as is standard in the $AdS/CFT$ correspondence, due to the warp factor
(see for example \vijay, and references therein).

From existing types of infinite extra dimension compactifications,
the model discussed here resembles most the examples of \visser,
and it is different from the more familiar infinite extra
dimension compactifications of the Randall-Sundrum (2) type
(\rubashap\akama\rstwo). In the compactification in this paper we
borrow only the time direction of $AdS$ to be part of the
``non-compact'' directions (and 4 dimensionality is achieved using
coordinates transverse to the $AdS$), whereas there all the
dimensions of space-time are from the $AdS$ directions.
Furthermore, in the RS2 type compactification one usually has a
brane and horizons at its two sides. The tension of the brane
usually has to be fine-tuned in order to have a 4 dimensional
Minkowski space, and therefore its stringy origin remains
unclear\foot{A realization of the RS scenario in String theory was
suggested in \hrmn. In \gkp\ it was shown that indeed one can
solve the hierarchy problem this way, but the issue of stability
of the background in this realization remains problematic.}. In
our case, the brane is replaced by the central region of global
$AdS_3$, where normalizable states localize, and the boundaries of
the space are the boundaries of $AdS$ itself (rather than
horizons). The model clearly exists within String theory (with the
addition of a full non-perturbative definition).

Another class of ``infinite extra dimension'' compactifications is one
in which the brane is embedded in a space which at infinity away from
the brane approaches flat space. One then tries to localize a graviton
on the brane (for example \infex\dvalib). In these models the
localization of the graviton is usually achieved by some fine tuning
at the region of the brane, whereas in our model the localization of
excitations is due to the behavior of space at infinity.

\subsec{Infinite extra dimensions and Spectrum Generating Algebras}

The reason that the supersymmetry breaking mechanism of \omeb\ does
not destabilize the background is the 1+1 dimensional conformal
symmetry of the background. A 3+1 observer will explain it as a result
of his world possessing a spectrum generating algebra. We would like
to argue that quite generally a spectrum generating algebra might be a
natural thing to explore in the context of infinite extra dimension
compactifications.

It is usually suggested that infinite extra dimension
compactifications might help solve the cosmological constant problem
by having corrections to the energy of a brane, or otherwise a loci
where excitations are trapped, change the geometry away from it,
rather than bending the brane itself\foot{The problem is then to
establish the existence of an effective 4D graviton, and to deal with
the continuum of states which often follows from the infinite extra
dimensions}. However, new algebraic structures might be another
important motivation, and in themselves might help to solve the
cosmological constant problem, or decouple states in some processes
etc. A non-trivial spectrum generating algebra is one example - ie,
the Hamiltonian of the theory is part of a more complicated
algebra\foot{Lorentz symmetry is of course such an algebra, but its
representation structure isn't rich enough to constrain the
cosmological constant.} which contains operators which do not commute
with it\foot{Although we will not require the entire spectrum to be in
a single representation of this algebra. Hence, it does not
``generate'' the entire spectrum}. In such cases the dynamics of
excitations at various energy scales would be related to each other,
and in particular their contributions to the cosmological constant
would be correlated. This might be a way towards solving the
cosmological constant problem.

Consider for example a warped compactification
\eqn\mtrc{ds^2=f(y)\eta_{\mu\nu}dx^\mu dx^\nu + g_{ij} dy^i dy^j}
where $\mu,\nu=0..3$ and $i,j=4,..9$ (if we want to restrict
ourselves to starting from 10 dimensions). Suppose the manifold
has a discrete symmetry which leaves the metric invariant but
rescales $f$ by some number $c$, i.e.
$$y=y({\hat y}),\ \ {\hat g}=g,\ \ \ {\hat f}(y)=cf(y).$$ Note that
there is no contradiction with Einstein's equations - if there is
a solution to Einstein's equations, then by rescaling $x$ there
exists a solution where $f$ is rescaled. A concrete example is the
Poincare patch of $AdS_p$, but we would like to consider the more
general case.

The action of a scalar field in this background is
$$\int d^4x d^6y \sqrt{g}f \biggl(
\eta^{\mu\nu}(\partial_\mu\phi)(\partial_\nu
\phi)+fg^{ij}\partial_i\phi\partial_j\phi+fV(\phi) \biggr)
$$

Under the transformation $y=y({\hat y})$, the action becomes
$$\int d^4x d^6y \sqrt{g} c f \biggl(
\eta^{\mu\nu}(\partial_\mu\phi)(\partial_\nu\phi) + c f g^{ij}
(\partial_i \phi)(\partial_j\phi)+c f V (\phi) \biggr)$$ If we append
this transformation by
$$x_{new}=c^{1\over 2} x$$ we return to the same Lagrangian. If the
3+1 dimensional theory does have a discrete spectrum of masses, this
transformation will have the effect of multiplying the masses of
particles by $c$. Hence warped compactifications have a chance of
giving rise to a string vacuum with interesting algebraic properties
in the mass spectrum.

It is very restrictive for a manifold to have such a symmetry, which
is how infinite extra dimensions come in - it is easy to show that no
such symmetry exists in the case that the internal manifold is
compact. In fact, in a Lorentz invariant theory if we assume that the
spectrum of masses is bounded from below by a finite gap above zero,
then we clearly cannot have a symmetry which acts by simply multiplying
the masses. However, more complicated algebraic structures are not
excluded, such as in the case that we are discussing.

This discussion was in a Lorentz invariant theory (and in non-Lorentz
invariant cases, one expects an even richer structure).  It is not
clear however whether spectrum generating algebras (other than Lorentz
or conformal symmetry) can be incorporated in a Lorentz invariant
theory at all, or in a theory on a $dS$ space (which is favored
observationally). But we can take an $AdS$ space and see whether we
can get a model with at least 3+1 large dimensions (although with no
Lorentz symmetry), which is our model.

\newsec{Double trace deformations and SUSY breaking}

The interest in a compactification of this type is twofold. The
first is that it is an infinite extra dimension compactification
with a discrete spectrum. The other is that, by extending the
framework of String theory to allow non-local string theories
\omea, it introduces new parameters which can be used to modify
the effective action. They do so by modifying the boundary
behavior of the theory, but due to the geometry of the
compactification, these modifications are translationally
invariant in $R^{1,3}$. In particular we can use such deformations
to break supersymmetry without destabilizing the vacuum.

Finding a stable non-supersymmetric vacuum is usually complicated in
String theory because, among other problems, tadpoles will be
generated for some fields, at some order in the genus expansion. In
the context of the AdS/CFT correspondence, tadpoles are dangerous only
for fields which correspond to marginal operators \misj\ (and a
tadpole can be generated only for a field which is invariant under all
the symmetries of the model). As was pointed out in \omeb, it is
therefore remarkable that one can find a SUSY breaking deformation of
the model which does not destabilize the vacuum. In the following we
will briefly review \omeb, and proceed to see how it will vary the 3+1
dimensional spectrum. In particular we will see that one can be in a
regime where the splitting between the fermionic and bosonic members
of the same former SUSY multiplet is arbitrarily large.

Of course, since we do not have Lorentz invariance, supersymmetry
is not 3+1 dimensional. Rather it closes only on time
translations. Furthermore, since we are in the NS sector of the
boundary theory, the supersymmetries are half integer moded, and
the $G_{\pm{1\over2}}$ do not commute with the Hamiltonian. When
we say that we break supersymmetry we mean that we break even this
algebra, and that the splitting of energies will be much larger
than the $\Lambda/2$ set by the structure of the NS-sector. At the
same time we can work in a regime where
$q>>\Lambda,\Lambda(1+n_L+n_R)$ in \disprel, where the dispersion
relation will be approximately Lorentz invariant\foot{This is not
much worse than most attempts at an infinite extra dimension where
there are typically many excitations, if not a continuum, below
the physically relevant range of momenta and masses.}.

Let us review some properties of the supersymmetry of the model.
The model has an $SU(2)_L\times SU(2)_R$ R-symmetry. We pick a
$U(1)_L\times U(1)_R$ subgroup of the R-symmetry group, which we
will denote as $J(z)$ and ${\tilde J}({\bar z})$, and we will view
the theory as a $(2,2)$ theory \gpnr\ ($z$ and $\bar z$ are
coordinates in the boundary CFT). We can bosonize the currents as
$J=i\sqrt{c/3}\partial\eta$, ${\tilde
J}=i\sqrt{c/3}{\bar\partial}{\bar\eta}$, where $c$ is the central
charge of the theory, which in our case is of order $k$ (see
\adstr), and $\eta$, ${\tilde\eta}$ are canonically normalized
scalar fields. We then decompose the operators in the theory as
\eqn\decomp{ O=e^{ip\eta+i{\tilde p}{\tilde\eta}}
P(\partial^n\eta,{\bar\partial}^m{\bar\eta}) {\hat O} } where $P$
is a polynomial and ${\hat O}$ is an operator in the CFT/$U(1)$.
Under this decomposition the supercharges have charges
\eqn\supcrg{ (p,{\tilde p})=(\pm \sqrt{3/c},0),\ \ \ (p,{\tilde
p})=(0,\pm {3/c}), } and we will denote $3/c$ by $\Delta$.  So far
this was general $(2,2)$ supersymmetry. In the specific case of
$AdS_3\times S^3\times T^4$, $c\sim k=Q_1Q_5$ and the spectrum of
charges \eqn\spctcrg{ p\sim {q\over \sqrt{c}}.\ \ \ {\tilde p}\sim
{{\tilde q}\over \sqrt{c}} } where $q$ and ${\tilde q}$ are
weights of $SU(2)$.

To visualize what follows one should think about the left and
right $U(1)$ symmetries as roughly arising from a single circle of
radius approximately $\sqrt{k}$, giving rise to the spacing of
charges above (although in the full theory we may not see all the
spectrum of the circle - we may need to do some additional
projections, and correlate the charges $(p,{\tilde p})$ with the
operator $\hat O$. But the analogy with the circle is enough for
the purposes of scaling). This is also clear in space-time - each
$U(1)$ acts as an isometry inside the $S^3$ whose radius is
proportional to $\sqrt{k}$). In this picture the supercharges have
both momentum and winding around this circle.

The susy breaking double trace \omeb\ deformation is then the
addition of \eqn\ssydef{ {\cal S}\rightarrow {\cal S}+{\tilde
h}\int d^2z J(z){\tilde J}({\bar z})} to the space-time CFT (and
on the GR side to boundary action) with arbitrary coefficient
$\tilde h$ (more details can be found in \omeb). This deformation
amounts to changing the radius of the $U(1)$ circle\foot{One
should emphasize that we are not squashing the $S^3$. That would
correspond to condensing on the boundary an operator that
corresponds to a metric field. Here we are condensing a double
trace operator which corresponds to a pair of gauge fields in the
bulk.}. If we have an operator with $(p,{\tilde p})$ then under
this deformation (and keeping $(\eta,{\tilde \eta})$ canonically
normalized) the momenta change as \eqn\chngmom{ \biggl( \matrix{p'
\cr {\tilde p}'} \biggr) = \biggl( \matrix{ cosh^2(\gamma) &
sinh^2(\gamma) \cr sinh^2(\gamma) & cosh^2(\gamma) } \biggr)
\biggl( \matrix{p \cr {\tilde p}} \biggr), } where $\gamma$ is a
function of $\tilde h$ ($\gamma=0\leftrightarrow \tilde h=0$).
Hence the dimensions of charged operators shift as we perform the
double trace deformation. In particular, the dimension of the
supercharges will shift, which means that we have broken
supersymmetry. Since this deformation can be shown to be truly
marginal we did not destabilize the backgroud.

We would like to see whether we get a large splitting in the 3+1
dimensional multiplets. The dimension of the operator $O$ \decomp\
receives a contribution from $e^{ip\eta+i{\tilde p}{\tilde\eta}}$
and from $P{\hat O}$. The total dimension is $1+\sqrt{1+m^2+q^2}$
(dimensions are corrected using $\Lambda$, and $q$ is the 3
dimensional momentum), where $m$ receives contributions both from
the mass in 10 dimensions and the angular dependence on $S^3\times
S^1$ (part of which are the $U(1)$ R-charges). By subtracting the
contribution of the exponential we deduce that the contribution of
$P{\hat O}$ is \eqn\dimho{1+n_L+n_R+\sqrt{1+m^2+q^2}-{
3(p^2+{\tilde p}^2)\over 2c }.} As we deform the theory, the
contribution from the exponentials in \decomp\ change. As a
function of $\gamma$ the total dimension is \eqn\totdim{
1+n_L+n_R+\sqrt{1+m^2+q^2}+{ 3(p^2(\gamma)+{\tilde
p}^2(\gamma)-p^2-{\tilde p}^2)\over 2c}, } where from \chngmom\
\eqn\newcont{ p^2(\gamma)+{\tilde p}^2(\gamma)={1\over 2}\biggl(
e^{ 2\gamma}(p+{\tilde p})^2+ e^{-2\gamma}(p-{\tilde p})^2 \biggr)
.}

Consider now taking $\gamma\rightarrow -\infty$. In this case, for
a state not to decouple from the spectrum - i.e., for its energy
in 3+1 dimensions not to go to infinity - it needs to satisfy
\eqn\notdec{ p={\tilde p}. } However, if this is true for some
given state, then it cannot be true for its partners in the
multiplet under an odd number of applications of the supercharges.
The latter will satisfy $p={\tilde p}\pm (2n+1) * \Delta$ for some
integer $n$, and hence will decouple from the 3+1 spectrum (the
dimension formula for a fermion is slightly different than
\disprel\ and \dimho, but the $(p,\tilde p)$ dependence is the
same). I.e, we concentrate on momentum modes along the circle of
the left and right $U(1)$ when we take its radius to infinity. The
supercharge have non-zero winding number in this convention and
hence so do some of the SUSY partners of the momentum modes. The
latter, however, are now lifted to infinity as we take the radius
of the circle to infinity. If the states that correspond to
momentum modes ($p=\tilde p$) are bosons then we certainly lift
all the fermions in the former multiplet to high energy (as well
as some of the bosons), and vice versa if the momentum modes are
fermions.

Furthermore, all of this can happen in a regime where $q>>m, 1,
p^2/c, {\tilde p}^2/c$, where the particle will have an
approximately Lorentz invariant dispersion relation.

\newsec{The 3+1 Dimensional Point of View}

In this section we will discuss what the quadratic action for a scalar
field on this space looks like, from the point of view of the 3+1
dimensional observer. We will begin with a scalar field of some mass
$m$ on $AdS_3\times T^3$, and KK reduce to $R\times T^3$. By the
action we mean the off-shell quadratic action for this field, which is
well defined since we are discussing a scalar field in a fixed
background. To compute the action we need to compute the eigenvalues
of the Laplacian on this space, and how they depend on the 3+1
momenta. This action that we will get will turn out to be a
generalized free field theory with some unusual features.

In general, in warped compactifications, one expects a generalized free
field theory already at the quadratic level when performing a
Kaluza-Klein reduction. Using the metric as in \mtrc\ we insert
$$\phi(x,y)=\Sigma_n \int d^4k e^{ikx} \phi_{n,k}(y)$$ where $x$ are
coordinates on $R^{1,3}$, $y$ are the rest of the coordinates and
$\phi_{n,k}$ is a complete set of functions of $y$ (for a given $k$)
which we will specify momentarily.  The action then becomes
\eqn\actna{\int d^4k \Sigma_n \int d^6y f^2 \sqrt{g}
\phi_n(y)^*\biggl( f^{-1} k^2\phi_n(y) + {1\over \sqrt{g} f^2}
\partial_i g^{ij} f^2\sqrt{g} \partial_j \phi_n(y) \biggr),} and we
choose $\phi_n$ to be eigenfunctions of the operator inside the
parenthesis (which is the full Laplacian on the space). The main
point is that the 4 dimensional momentum, $k$, appears explicitly
in the eigenfunctions and eigenvalues on the 6 dimensional
manifold - the space-time momentum dependence of the action need
no longer be quadratic\foot{Some non-quadratic actions are
intimately related to new symmetries. The familiar example is that
of the DBI action \rob\bgrglp.} in momentum once we go to the
basis of functions $\phi_{n,k}$.

The off-shell states in the low energy effective action can be
derived by a KK reduction, which we will discuss in a minute.
However, since the conformal field theory on $SL(2,R)$ was
extensively studied over the last few years we can guess the
results. Generally in String theory the off-shell effective action
is not well defined, since String theory allows only the
computations of on-shell quantities. But it does provide a natural
guess as to the off-shell modes of some fields in some cases. In
the large $M_s$ limit, relaxing the equation of motion in
space-time is equivalent to relaxing the constraint that
$L_0+{\tilde L}_0=2$ on the operators in String theory. Hence as
off-shell states we will take all operators in the conformal field
theory that satisfy all physical state conditions, with the
exception that we relax this condition.

The spectrum of relevant representations is given in \maoga. We will
focus on the spectrum of unflowed representations since we will work
in the point-particle limit, below the ``long strings'' threshold
\mlstlong\sewtlong. We will denote the generators of $SL(2,R)$ by
$J^{\pm}$ and $J^3$, where $J^3$ corresponds to time translations (we
are working with global $AdS_3$ and hence its spectrum is
continuous). There are 2 types of relevant representations of
$SL(2,R)\times SL(2,R)$:

\medskip

\item{1.} A product of principal discrete representation for the
left and right movers ${\cal D}^\pm_j \times {\cal D}^\pm_j$ with
$j>{1\over 2}$. The spectrum of these representations satisfies
\eqn\dscrtd{\pm w=1+\sqrt{1+q^2+m^2-\lambda}+n_l+n_r,} where
$\lambda$ is the eigenvalue of the total Laplacian (with the mass
term) on $AdS_3\times T^3$, and we are interested as $\lambda$ as
function of $w$ and $q$. In this notation
$j={1\over2}(1+\sqrt{1+p^2+m^2-\lambda^2})$.

\item{2.} A product of principal continuous representations
${\cal C}^{\alpha}_{j={1\over2}+is}\times {\cal C}^{\alpha}_{j={1\over
2}+is}$.
In these representations the eigenvalues of the $AdS_3$ Laplacian and
$J_3$ are uncorrelated. We will mix the representations by writing all
of them together as $|j,w_l,w_r\rangle$, $w_l,w_r>0$,
$j={1\over2}+is$. These representations do not contribute to on-shell
degrees of freedom.

\medskip

The quadratic low energy effective action now becomes
\eqn\lwenrg{S_{off-shell}=S_{off-shell,discrete}+S_{off-shell,continuous}}
where \eqn\lwenrga{ S_{off-shell,discrete}=\Sigma_{n_l,n_r\ge
0}\int_{|w|>1+n_l+n_r} dw \int d^3q }
$$\phi^*_{w,q,n_l,n_r}
\bigl( (|w|-1-n_l-n_r)^2-q^2-m^2 \bigr) \phi_{w,q,n_l,n_r}
$$ and $S_{off-shell,cont.}$ is a higher dimensional integral over
$w,k$ and the additional continuous parameters in item 2 above. The
details of the latter do not matter for us, except that there is no zero
eigenvalue of the Laplacian in this sector.

The three interesting features that we see in this off-shell
action are: \item{1.}  The discrete states are located on lines
$\lambda(w)$ (for fixed $q$) satisfying \dscrtd. A 3+1 observer
will associate such a line with the off-shell states of a single
3+1 particle (with the mass shell condition being $\lambda=0$).
Such an observer will report that we can define the off-shell
modes of a particle only above a certain frequency. Hence we do
not have off-shell Green's functions for the 3+1 fields for
arbitrary position or momentum. In particular the zero frequency
mode was removed, even in the off-shell action. \item{2.} The
action is non-analytic in the frequency $w$. Correspondingly, when
Fourier transforming the action for the discrete modes into
position space we obtain a non-local action. \item{3.} A continuum
appears, with no physical degrees of freedom\foot{A perhaps more
familiar appearance of a path integral with no physical degrees of
freedom is in open String field theory after the condensation of
the tachyon. There one still writes an action for the open string,
but there are no open string physical degrees of freedom.}.

It is straightforward to obtain the same results by an ordinary
Kaluza-Klein reduction. This is nothing but an exercise in
expanding the functions on $AdS_3\times T^3$ in a complete set of
states (the two sets of representations are a complete set of
states for the ordinary $L^2$ norm). A discussion of the discrete
representations essentially appears in \vijay\ (and references
therein), with the slight modification that there one was looking
for on-shell states (and also $p^2$ was not differentiated from
$m^2$), and hence one sets $\lambda=0$. This gives a relation
between $m^2$ and $w$. However, since a non-zero $\lambda$ is the
same as a different $m$, to go to our case one simply needs to
replace $m^2$ there by $p^2+m^2-\lambda$.

It is also easy to see where the continuous representations come from. We
will follow the notation in \vijay\ where the metric is
$$ds^2={-dt^2+d\rho^2\over cos^2(\rho)} + tan^2(\rho)d\theta^2$$ and
the boundary of $AdS$ is at $\rho=\pi$, and we will denote
$z=cos(\rho)$. To have a well-posed problem we will work with a finite
cut-off at some finite small value of $z=\epsilon$. At the end of the
computations we will take $\epsilon$ to 0. We need to impose some
boundary condition at $z=\epsilon$. The details of the boundary
condition will not be important, but for concreteness we can use the
boundary conditions in \bss\ $(z\partial_z-1-\sqrt{1+q^2+m^2})\Phi
\vert_{z=\epsilon}=0$.  Denoting $\nu=\sqrt{1+q^2+m^2-\lambda}$, the
small $z$ behavior of the wave function is \vijay \eqn\cont{
{\Gamma(-\nu)\over\Gamma\bigl({1\over2}(1-\nu+l+w)\bigr)\Gamma\bigl({1\over2}(1-\nu+l-w)\bigr)}*}
$$*z^{1+\nu}F\biggl({1\over2}(1+\nu+l+w),{1\over2}(1-\nu+l-w);1+\nu;z^2\biggr)+$$
$$+{\Gamma( \nu)\over\Gamma\bigl({1\over2}(1+\nu+l+w)\bigr)\Gamma\bigl({1\over2}(1+\nu+l-w)\bigr)}*$$
$$*z^{1-\nu}F\bigl({1\over2}(1-\nu+l+w),{1\over2}(1-\nu+l-w);1+\nu;z^2\bigr).$$
To satisfy the boundary condition we need to balance the 2 terms,
which requires the 2 terms to be the same order of magnitude.  For
fixed real $\nu$, the 1st term nominally decays much faster than the 2nd
term when $z\rightarrow 0$, and to remedy this we require that one of
the $\Gamma$ functions in the denominator of the 2nd term diverges -
this gives us the discrete representations. To obtain the continuous
representation we take $\nu$ to be imaginary, which gives us another
way of cancelling the 2 terms. For finite $\epsilon$ the spectrum
would be discrete but it would go to a continuum as
$\epsilon\rightarrow 0$.

To conclude this section we would like to comment on the relation
between the fact that the model is non-Lorentz invariant and the fact
that it has a spectrum generating algebra. The dispersion relation
$$w_{n,q}=f(n_L,n_R)+\sqrt{1+q^2+m^2}$$ is an interesting way in which
one can combine the two, where the algebra changes $n_L$ and
$n_R$. The velocity of a wave packet $\partial w_{n,q}/\partial q$ is the same for all the
particles in the multiplet. Hence we can have a locally acting
spectrum generating algebra at the price of introducing non-lorentz
invariant dispersions relations of the form \disprel.

\newsec{Generalizations}

{\bf Uniqueness of the Basic Model}

We have discussed so far the background $AdS_3\times S^3\times
T^4$. We would like to know whether there are other backgrounds to
which we can apply our SUSY breaking mechanism. As a first step we
would like to consider all backgrounds of the type $AdS_3\times
T^3\times {\cal N}$, where ${\cal N}$ is a compact conformal field
theory, such that the space-time theory has at least $(2,2)$
supersymmetry.  We will show that under these circumstances the
model actually has $(4,4)$ supersymmetry. This leaves us with the
$AdS_3\times S_3\times T^4$ as the main model\foot{There are other
models that have $(4,4)$ supersymmetry, but these reduce to the
model above locally, ie., when we go to the regime of parameter
space where 3 coordinates become large (in addition to time). For
example, $AdS_4\times S^3\times K3$.}, and models built on this
basic example, which we will mention later.

We will use \susycond, which specifies the conditions under which
$AdS_3\times {\cal N}'$ will have $(2,2)$ supersymmetry. As there
$SL(2)$ will be taken at level $k$, leading to central charge
$c^{SL(2)}={3(k+2)\over k}+{3\over 2}$. To have supersymmetry we
require that:

\item{1.} ${\cal N}$ contains an affine $U(1)$ current
$\psi^{U(1)}+\theta J^{U(1)}$

\item{2.} ${\cal N}/U(1)$ is an $N=2$ superconformal theory with central
charge $c^{{\cal N}/U(1)}={9/2}-6/k$. We will bosonize its current
with a canonically normalized scalar $J^{{\cal N}/U(1)}_R=ia\partial
z$, where $a=\sqrt{c^{{\cal N}/U(1)}/3}$.

The $SL(2)$ is made out of 3 bosonic $SL(2)$ current and 3 free
fermions $\Psi^{1,2,3}$.
We will define the following bosons
$$\partial H_1=\Psi^1\Psi^2$$
$$\partial H_2=\Psi^3\Psi^{U(1)}$$
$$i\sqrt{3}\partial H_0= J^{{\cal N}/U(1)}-\sqrt{{2\over k}}J^{U(1)}$$
The space-time susy generators are
$$G_r^\pm\propto \int dz e^{-\phi/2} S_r^\pm {\dot S}, \ \ \ r=\pm
{1\over 2}$$ where $$S_r^\pm=e^{ir(H_1\pm H_2)\pm i(\sqrt{3}/2)H_0}$$
and ${\dot S}$ are the spin operators from the $T^3$ directions. We
end up with the model having at least 2 complex supersymmetries in
each of the left and right moving sector. This implies that the model
has at least $(4,4)$ supersymmetry.

{\bf More Examples}

Once we have constructed the basic example as above, we can modify it
in different ways to generate a richer spectrum. Again we would like
to concentrate on examples in which the SUSY breaking mechanism
described above is still valid. This means that we can do any
modification as long as we do not change the boundary CFT, ie, we can
do any change that we want in the interior of $AdS_3$, which would
correspond to turning on states in the CFT, but we cannot deform the
behavior near the boundary. This means that we are still discussing
the same theory which has the SUSY breaking deformation.

Two kinds of deformations are potentially interesting as they may give
solvable CFTs. One is orbifolding the $AdS_3\times S^3\times T^4$ in a
way which does not act on the time or on the large $T^3$. Examples of
these were discussed in \emila\emilb\vijcon. Another modification is
to place some branes at the origin of $AdS$, wrapping the $T^3$. The
fact that the branes do not intersect the boundary of $AdS_3$ means
that they correspond to states in the same CFT as before.

These branes can have various configurations on the $S^3$ and on the
remaining $S^1$. If one generalizes the model both by orbifolding and
adding branes to the orbifold, a very rich set of spectra can emerge.
Relative to ordinary compactifications we can add branes much more
liberally. The reason is that charge conservation is less of an
obstacle here since the manifold is topologically non compact and the
flux can escape to infinity.

This might also help in terms of the Lorentz invariance of the
spectrum. Consider the simplest case of a brane wrapping the time
direction and the $T^3$ and situated at the origin of $AdS_3$. The
metric that excitations on this brane see is
$$-g_{00}(\rho=0)dt^2+dx_i^2,$$ which has a Lorentz symmetry. This
symmetry is, of course, not exact because of coupling to non-lorentz
invariant background fields, non-lorentz invariant closed strings
etc. Still, one expects that Lorentz invariance will be improved in
the D-brane sector vs. the closed string sector because the low energy
closed strings are states with width of order $1/\Lambda$ on the
$AdS$, whereas low lying open strings are of width $l_s$ around the
position of the brane. If we take $l_s<1/\Lambda$, as we have been
doing so far, the smaller width on the $AdS$ will translate into
improved Lorentz invariance.

\vskip 1cm

\centerline{\bf Acknowledgements}

It is a pleasure to thank O. Aharony, S. Elitzur, B. Fiol, A. Giveon,
D. Kutasov, E. Rabinovici, M. Rozali, A. Schwimmer, A. Sever,
Y. Shadmi, A. Shomer, J. Simon and E. Silverstein for useful discussions. This
work was supported by the Israel-U.S. Binational Science
Foundation, the IRF Centers of Excellence program, the European RTN
network HPRN-CT-2000-00122, and by the Minerva foundation.

\listrefs

\end